# Magnetic Cellular Nonlinear Network with Spin Wave Bus for Image Processing

Alexander Khitun, Mingqiang Bao, and Kang L. Wang

Device Research Laboratory, Electrical Engineering Department,

University of California, Los Angeles

FCRP Focus Center on Functional Engineered Nano Architectonics (FENA),

Nanoelectronics Research Initiative - The Western Institute of Nanoelectronics (WIN).

## Abstract

We describe and analyze a cellular nonlinear network based on magnetic nanostructures for image processing. The network consists of magneto-electric cells integrated onto a common ferromagnetic film – spin wave bus. The magneto-electric cell is an artificial two-phase multiferroic structure comprising piezoelectric and ferromagnetic materials. A bit of information is assigned to the cell's magnetic polarization, which can be controlled by the applied voltage. The information exchange among the cells is via the spin waves propagating in the spin wave bus. Each cell changes its state as a combined effect of two: the magneto-electric coupling and the interaction with the spin waves. The distinct feature of the network with spin wave bus is the ability to control the inter-cell communication by an external global parameter - magnetic field. The latter makes possible to realize different image processing functions on the same template without rewiring or reconfiguration. We present the results of numerical simulations illustrating image filtering, erosion, dilation, horizontal and vertical line detection, inversion and edge detection accomplished on one template by the proper choice of the strength and direction of the external magnetic field. We also present numerical assets on the major network parameters such as cell density, power dissipation and functional throughput, and compare them with the parameters projected for other nano-architectures such as CMOL-CrossNet, Quantum-Dot Cellular Automata, and Quantum Dot-Image Processor. Potentially, the utilization of spin wave phenomena at the nanometer scale may provide a route to low-power consuming and functional logic circuits for special task data processing.

**Introduction**

Image processing is a well-established technique required for a variety of practical applications. Image processing on a general-purpose computer is a time and resources consuming task. The basic image processing functions such as filtering require a number of consequent pixel-by-pixel operations for the entire image. The processing time increases proportional to the size of the image. In spite of the rapid progress in the computers functional throughput enhancement, the real-time image processing of a large size images is still a challenging problem. It would be of great practical benefit to develop special task data processing hardware and integrate it on silicon platform as a complimentary logic unit to a general-purpose processor. The distinct feature of such a device would be the ability to process a number of bits (pixels) in parallel at constant time and to perform the specific logic functions required for image processing.

Cellular Nonlinear Network (CNN) has been received a growing deal of interest as a promising architecture for parallel data processing and future computation using nanometer scale devices and structure. The concept of CNN was first formulated by Leon Chua [1]. CNN is as a 2 (3 or more) dimensional array of mainly identical dynamical systems, called cells, which satisfy two properties: (i) most interactions are local within a finite radius $R$, and (ii) all state variables are signals of continuous values. The cell is a nonlinear dynamic system, which behavior is governed by the following nonlinear ordinary differential equation:

$$C\frac{dx_{ij}(t)}{dt} = -\frac{x_{ij}(t)}{R} + \sum_{C(k,l)\in N_r(i,j)}(a_{ij,kl})f(x_{kl}(t)) + b_{ij,kl}u_{kl}) + I \quad (1)$$

$$1 < i < M; \quad 1 < j < N; \quad f(x_{ij}) = 0.5 \times (\|x_{ij} + 1\| - \|x_{ij} - 1\|)$$

where $N(i,j)_r$ represents the neighborhood of cell $(i,j)$ in an $M \times N$ CNN array; $x_{ij}(t)$, $u_{ij}$, and $f(x_{ij})$ are state, input, and output variables of cell$(i,j)$, respectively. The space invariant feedback and feed-forward parameters $a_{ij,kl}$ and $b_{ij,kl}$ are showing the weighting for the feedback and feed-forward interaction among cell$(k,l)$ and cell $(i,j)$. $I$ is the cell bias. In the series of works, the CNN paradigm was evolved in many ways and powerful computing abilities of the CNN, especially for image processing, were demonstrated by numerical modeling[2-5].

Potentially, molecules, atoms or spin systems can be used as cells connected in a network. Such networks built of nanocells would have a huge integration density and may benefit from the utilization of quantum –mechanical phenomena such as tunneling or exchange interaction for inter-cell interconnection. On the other hand, there are certain technological constrains which may significantly limit the functional throughput of the nano-CNN. The most challenging technical problems on the way towards nano-CNN practical implementation are associated with: (i) nano-cell size variability; (ii) difficulty in providing individual nano-cell addressing, and (iii) control of the inter-cell interaction. The inevitable variations of the nano-cell size and position will result in additional power dissipation and might need error-correcting logic units. The lack of individual cell addressing or/and inter-cell interaction control would significantly restrict the number of logic functions, which can be realized on one nano-template. From the mathematical point of view, in order to implement one or another image processing function, one has to change the feedback and feed-forward parameters in Eq.(1). Physically, it implies the control of the individual cell parameters and/or the change of the inter-cell interaction. It is tremendously difficult to provide an individual wiring for a large number of nano-cells. It is even more challenging problem to control/change selectively the interaction between the nano-cells in a dense array. One of the possible approaches to this problem is to exploit non-linear interactions among the nano-cells. In this case, different logic functions can be accomplished by varying the boundary conditions. The examples of CNN with a non-linear inter-cell coupling are given in [6-8]. Another possible approach is to use an external field as a global parameter, which can modulate the inter-cell interaction.

In this work, we describe a magnetic CNN (MCNN), where the interaction among the cells is controlled by an external magnetic field. There is an impetus to the development of nanometer scale magnetic materials and structures for novel computing devices. The potential advantages of using magnetic or spin-based materials are twofold: (i) the magnetization of a single domain or nanostructure can be used as a non-volatile memory cell, and (ii) the interaction among the magnetic cells can be performed without the use of electric currents, which may be a route to low-power consuming architectures. Among the proposed spin-based architectures, there are Magnetic Cellular Automata (MCA) [9], Domain-Wall Logic[10], and networks with Spin Wave Bus [11]. In the previous works [12-15], we have studied possible spin wave based logic devices and developed the concept of magnetic NanoFabric for general and special task data processing. Magnetic NanoFabric is the combination of electrically-controlled magnetic elements (e.g. multiferroics) with ferromagnetic waveguides for spin wave propagation. The advantage of this approach is that the wave nature of the spin wave phenomena allows us to utilize wave superposition and interference for achieving parallel data processing. In the next section, we describe the material structure and the principle of operation of the magnetic CNN with Spin Wave Bus. In Section III, we present the results of numerical simulation illustrating network performance. Then, we evaluate and discuss the major CNN parameters such as cell density, power consumption and functional throughput in Section IV. We also discuss the advantages and shortcomings of MCNN in comparison with other proposed nano-architectures for image processing. Conclusions are given in Section V.

II. Material Structure and Principle of Operation

The schematic of MCNN in conjunction with two electronic circuits is shown in Fig.1. One electronic circuit is to provide the input data (e.g. an image to be processed) and the second circuit is for MCNN programming (to define a specific processing function). The input data are received from the electronic circuit in the conventional form of voltage pulses. Then, the voltage pulses are translated into the spin wave signals by the converters shown at the edge of the MCNN. Information exchange and computations within MCNN are accomplished via the spin waves. The result of data processing is converted to the voltage pulses and returned to the electronic circuit. The programming electronic circuit is aimed to control the external magnetic field affecting spin wave propagation within MCNN.

The core of the MCNN is a layered structure comprising from the bottom to the top a semiconductor substrate (e.g. silicon), a conducting ferromagnetic layer (e.g. NiFe), an insulating layer (e.g. silicon dioxide) with a regular two dimensional array of nano-patterns filled with a non-conducting piezoelectric material (e.g. PZT), and a metallic contact (e.g. Al) on top. The structure of the electro-magnetic cell is shown on the right in Fig.1. The cell consists of the ferromagnetic material under the piezoelectric nano-pattern, piezoelectric material, and the metallic contact on the top. The conducting ferromagnetic layer and the top electrode serve as two sides of the parallel plate capacitor filled with the piezoelectric. An electric field applied across the piezoelectric layer produces stress. In turn, the stress results in the rotation of the easy axis in the ferromagnetic material under the piezoelectric. Such a cell comprising piezoelectric and ferromagnetic materials represents a two-phase multiferroic element [16]. An electric field applied across such a structure changes the magnetic polarization in the ferromagnetic material and vice versa. The strength of the coupling is described by the magnetoelectric coefficient in V/(cm Oe). The advantage of using a two-phase multiferroic is that each material may be independently optimized to provide prominent electromagnetic coupling. There are several piezoelectric-ferromagnetic pairs, which have been experimentally studied, showing a prominent magneto-electric coupling: PZT/NiFe$_2$O$_4$ (1,400 mV cm$^{-1}$ Oe$^{-1}$)[16], CoFe$_2$O$_4$/BaTiO$_3$ (50 mV cm$^{-1}$ Oe$^{-1}$)[17], PZT/Terfenol-D (4,800 mV cm$^{-1}$ Oe$^{-1}$)[18].

All ME cells in the array are sharing the same ferromagnetic layer. The latter makes possible to accomplish the interaction among the cells via the spin waves propagating in the ferromagnetic film –spin wave bus. In the preceding work [15], we described the idea of using a continuous ferromagnetic film as a conduit for spin wave propagation and as a host material for ME cells. It was suggested to use a crossbar structure consisting of piezoelectric rows and metallic columns to form an array of ME cells in the region under the intersection between the rows and columns. In this work, we further simplify the structure by replacing the crossbar structure by the nano-pattern layer. None of the ME cells has an individual contact, or a bias wire, but all cells are served by two common electrodes: the conducting ferromagnetic layer on the bottom and the metallic contact on the top. The latter makes the proposed structure attractive from the fabrication point of view.

A bit of information in MCNN is encoded into the in-plane magnetic polarization (parallel to the ferromagnetic layer surface) of the ME cells. We assign ME cell to logic state 0 if it has a magnetic polarization component along the X-axis ($M_x>0$). We assign ME cell to logic state 1, if the cell has magnetic polarization opposite to the X-axis ($M_x<0$). It should be clarified that the cells may have out of plane (perpendicular to the ferromagnetic layer surface) magnetic polarization much higher than the in-plane ($M_z>>M_x$) during the operation. However, only one magnetization projection $M_x$ is used for information storage. The information exchange among the ME cells is via spin waves propagating in the spin wave bus. A spin wave is a collective oscillation of spins in ferromagnetic material, which can be excited by the external magnetic field (e.g. by the ME cells as a result of the magneto-electric coupling). The operation of the ME cell has been illustrated by numerical modeling in [15]. Here, we would like to explain briefly the basic of the ME operation. If there is no bias voltage applied to the ME cell, we assume that the easy axis of the ferromagnetic layer is directed perpendicular to the surface of the ferromagnetic (along the Z axis) as shown in Fig.2(a). In this state, the cell has no magnetic polarization component along or opposite to the X axis. By applying voltage to the ME cell, the easy axis is rotated towards the in-plane direction (from the Z axis towards the X axis). The rotation of the easy axis makes two directions (along or opposite to the new easy axis) energetically preferable as shown in Fig.2(b). The higher is the rotation angle, the bigger is the magnetization projection on the X axis, and the higher is the energetic barrier ΔE between the 0 and 1 states. In the ultimate limit of 90 degree easy axis rotation shown in Fig.2(c), the cell has the biggest $M_x$ component and the highest energetic barrier ΔE. At the moment of the easy-axis rotation, a small perturbation caused by a spin wave(s) can define the final direction of the cell magnetization. The switching has a threshold-like behavior. To switch the cell magnetic polarization, the exchange field produced by the spin waves should be higher than the coercive field to overcome ΔE. The threshold value is defined by the material parameters and the strength of the electro-magnetic coupling. Lowering or increasing the applied electric field, we assume to control the threshold and to make the cell switchable or immune to the incoming spin waves.

The communication between the MCNN and the electronic circuits is via the converters - the devices translating voltage pulses into the spin waves and vice versa. An example of the converter is a simple conducting loop placed in the vicinity of the ferromagnetic film. An electric current in the loop produces magnetic field, which excites spin waves in the ferromagnetic media. The direction of spin precession in the ferromagnetic film (the phase of the excited spin wave) depends on the direction of the electric current in the loop. This fact has been used in the prototype spin wave logic device described in [19]. The detection of the spin wave by the conducting loop is via the inductive voltage measurement. A propagating spin wave changes the magnetic flux through the conducting contour. According to Faraday's law, the change of the magnetic flux induces an inductive voltage in a conducting loop, which magnitude is proportional to the rate of the magnetic flux change $E_{ind} = -d\Phi_m/dt$. For high frequency operation, the converter can be made in the form of a microstrip or an **a**symmetric **co**planar **s**trip (ACPS) transmission line [20-22]. In this work, we consider just two ACPS line as the input/output ports for the whole MCNN providing spin wave excitation and detection as shown in Fig.1.

The operation of the MCNN includes three major steps: information read-in (ME cell initial polarization), data processing (spin wave exchange among the ME cells), and information read-out (the detection of the final state of each ME). These steps are illustrated in figures 3-5. In order to simplify the explanation, we make the following assumptions. The easy axis of the ferromagnetic layer is parallel to the Z axis. The applying of the bias voltage locally rotates the easy axis in the ferromagnetic under the piezoelectric pattern from the Z axis towards the X axis. We restrict our consideration by the surface magnetostatic spin waves propagating on the surface of the ferromagnetic layer. The group velocities in the X and Y directions are the same unless the bias magnetic field is tilted toward the X or Y axes.

The read-in procedure is accomplished by using the interference of two spin waves generated by the ACPS lines as shown in Fig.3(a). Two ACPS lines generate coherent plane waves propagating along the X and Y axes. At the point of wave intersession, the amplitude of the local magnetization change $\Delta M_x$ can be twice the amplitude produced by a single spin wave (constructive interference). The point of intersection moves in the X-Y plane as it is illustrated in Fig.3(b). In order to read-in a particular cell in the array, the bias voltage is decreased at the moment when the point of interference crosses the cell. The read-in voltage is adjusted in such a way that the threshold value of ME switching $M_{th}$ is higher than the amplitude of the single plane wave but less than the double amplitude $\Delta M_x < M_{th} < 2\Delta M_x$. In Fig.3(c), we show three snapshots illustrating the propagation of two plane waves through the ferromagnetic layer. The time is depicted in the normalized units of $t_0$, which is the time spin wave travels across a single ME cell. For example, in the moment in time $t=10t_0$, the point of waves intersection crosses the cell with coordinates 30,30 in the XY plane. The bias voltage is lowered (threshold lowered) at this moment of time to allow cell switching. Then, the bias voltage is increased to the previous value making all cells in the array insensitive (not-switchable). The bias can be decreased again to allow next cell switching (e.g. cell 60,60 as it is shown in Fig.3(b)). The coordinate of the wave intersection point $(x,y)$ is defined by the spin wave group velocities in the perpendicular directions $v_x$, and $v_y$ and the time the wave being excited $t_x$ and $t_y$ ($x = v_x \cdot t_x$, $y = v_y \cdot t_y$). Every cell in the array can be addressed by introducing the time delay between the ACPS lines $(t_x, t_y)$. The reasonable question to ask is how fast one has to change the bias voltage to resolve in time the nearest neighbor cells? For typical spin wave velocity of $10^5$cm/s -$10^7$cm/s, the time delay between two adjacent cells separated by 100nm distance is about 0.1ns-1ps, which is in the frequency range of today's electronic circuitry.

Computation in MCNN is associated with the change of the ME cell magnetic polarization as a function of the polarization of its neighbor cells. It is accomplished in a parallel manner for all cells in the network via the spin wave exchange. At the beginning of the computational step, the bias voltage is lowered. The decrease of the bias voltage affects the anisotropy field (easy-axis rotation from X axis towards the Z axis) due to the electro-magnetic coupling. As a result, each cell emits spin waves. In Fig.4, we show the waves emitted by the cells as the cylindrical waves propagating in all directions through the ferromagnetic layer. The amplitude of the emitted spin waves is the same of all cells. However, the phase of the wave depends on the initial cell polarization along or opposite to the X-axis. After the time required for spin wave propagation to the nearest-neighbor cells, the bias voltage is increased. The easy axis for each cell is rotated from the Z axis towards the X axis. The final magnetic polarization along or opposite to the X axis for each cell is now defined by the sum of all incoming spin waves and the wave produced by the cell itself. As the spin wave propagation is sensitive to the external magnetic field [20], we propose to exploit this fact to control the interaction between the ME cells. The phases of the incoming waves can be manipulated by lowering or increasing the strength of the external magnetic field. For example, the spin wave produced by a ME cell polarized along the X-axis may arrive with 0 or a π-phase shift to the nearest-neighbor cell. This manipulation with the phase is equivalent to the buffer or inverter function providing the same or the inverted signal from one cell to another. In the next section, we present the results of numerical modeling illustrating possible useful logic functionality exploiting spin wave phase modulation by the external magnetic field. After one or several steps of computation, the magnetization of the ME array evolve to some configuration – the result of the data processing.

There are at least two possible mechanisms for read-out procedure. First, it can be done in the similar manner as the read-in procedure. Each cell in the array can be address by the superposition of two plane waves generated by ACPS lines (as it is shown in Fig.5). The same ACPS lines can be used to detect the backscattered signal from the cell. The information on the cell state ($M_x>0$ or $M_x<0$) is reflected into the phase of the scattered wave (0 or $\pi$ initial phase). The sign of the produced inductive voltage shows the phase of the detected spin wave. Experimental data on the time-resolved spin wave detection via the inductive voltage measurements can be found elsewhere [20, 21]. The described read-out mechanism resembles the well-known method of radiolocation [23], but applied at the nanometer scale. Another possible mechanism for read-out is to excite all cells simultaneously by lowering the bias voltage. As a result, all cells in the array will emit spin waves. The emitted spin waves then can be detected by the ACPS lines. In this scenario, each line will detect the signals not from a single cell but from a number of cells. As the spin waves from the distant cells will arrive with a certain time delay, it is possible to extract the information on the individual cell. The second mechanism with simultaneous cell excitation may provide a faster read-in procedure but it would require an additional logic circuit for the retrieved data decoding.

### III. Numerical Modeling

The dynamics of MCNN can be modeled using the Landau-Lifshitz equation:

$$\frac{d\vec{m}}{dt} = -\frac{\gamma}{1+\alpha^2} \vec{m} \times \left[\vec{H}_{eff} + \alpha \vec{m} \times \vec{H}_{eff}\right], \qquad (2)$$

where $\vec{m} = \vec{M}/M_s$ is the unit magnetization vector, $M_s$ is the saturation magnetization, $\gamma$ is the gyro-magnetic ratio, and $\alpha$ is the phenomenological Gilbert damping coefficient. The first term of the right hand site of Eq.(2) describes the precession of magnetization about the effective field and the second term describes its dissipation. The effective magnetic field $\vec{H}_{eff}$ is given as follows:

$$\vec{H}_{eff} = \vec{H}_d + \vec{H}_{ex} + \vec{H}_a + \vec{H}_b, \tag{3}$$

where $\vec{H}_d$ is the magnetostatic field ($\vec{H}_d = -\nabla \Phi$, $\nabla^2 \Phi = 4\pi M_s \nabla \cdot \vec{m}$), $\vec{H}_{ex}$ is the exchange field ($\vec{H}_{ex} = (2A/M_S)\nabla^2 \vec{m}$, $A$ is the exchange constant), $\vec{H}_a$ is the anisotropy field ($\vec{H}_a = (2K/M_S)(\vec{m} \cdot \vec{c})\vec{c}$, $K$ is the uniaxial anisotropy constant, and $\vec{c}$ is the unit vector along the uniaxial direction), $\vec{H}_b$ is the external bias magnetic field. The magneto-electric coupling can be taking into account by introducing a voltage dependent anisotropy field. In our consideration, the unit vector $\vec{c}$ is directed along the Z-axis in the absence of the external electric field. The application of the bias voltage $V$ to the ME cell results in the easy-axis rotation ($\vec{c}$ vector rotation). For simplicity, we assume the components of the unit vector $\vec{c}$ to be the functions of the applied voltage as follows:

$$c_x = c_0 \sin\theta, \; c_y = 0, \; c_z = c_0 \cos\theta, \tag{4}$$

$$\theta = \frac{\pi}{2}\left(\frac{V}{V_G}\right),$$

where $V_G$ is the voltage resulting in a 90 degree easy axis rotation from the Z axis towards the X axis. The voltage-dependent anisotropy field appears only for the ME cell as a result of the electro-mechanical-magnetic coupling. Rigorously, the anisotropy field is a function of the stress produced by the piezoelectric under the applied voltage. The anisotropy field cannot be changed everywhere under the top electrode, but only in the regions under the piezoelectric.

The set of equations (2-4) can be used to simulate the processes of information transmission and processing in MCNN including spin wave excitation by ACPS lines, spin wave propagation through the ferromagnetic layer, ME cell switching, spin wave excitation by ME cells. Some of this processes have been modeled in our preceding works [15, 24]. In Fig. 6, we present the results of numerical simulations showing the magnetization dynamics of a single ME cell as a result of the electro-magnetic coupling and interaction with a spin wave. At the initial moment of time, the cell is polarized along the Z-axis ($M_x = M_y = 0$, $M_z = 1$). There is no bias voltage applied, and the easy axis is parallel to the Z direction. Then, an incoming spin wave arrives to the ME cell (e.g. spin wave may be excited by the nearest neighbor cell). At the moment of the spin wave arrival (the wave front reaches the region under the PZT), the bias voltage is applied initiating the easy-axis rotation towards the X-axis. The perturbation caused by the spin wave $\Delta M_x$ is much smaller than the saturation magnetization $M_s$. However this small perturbation defines the trajectory of the cell evolution along or opposite the X axis. In Fig.6, there are shown two evolution curves: the solid curve corresponds to the case when the incoming spin wave has a positive x-component ($\Delta M_x > 0$), and the dashed curve corresponds to the case when the incoming spin wave has a negative x-component ($\Delta M_x < 0$). In our numerical simulations, we used $\gamma = 2.0 \times 10^7$ rad/s/Oe, $4\pi M_s = 10$kG, $2K/M_s = 4$Oe, $H_b = 50$Oe, and $\alpha = 0.1$. The gyro-magnetic ratio, saturation magnetization, and the anisotropy constant are the typical values for permalloy films [20, 25]. The Gilbert coefficient is intentionally increased to show rapid relaxation. We did not use any particular value of the bias voltage $V_G$ but assumed the final easy-axis direction is along the X-axis. The results shown in Fig.6 illustrate the ME cell switching as a combined effect of the interaction with the spin wave and the magneto-electric coupling. The cell becomes polarized along the X axis if $\Delta M_x > 0$, and opposite to the X axis if $\Delta M_x < 0$. We use this result as a starting point for the subsequent speculations on the MCNN logic functionality. Following the described procedure, the cell can be switched not only by a single wave but the superposition of a number of waves. The final cell polarization is defined by the sum of the amplitudes of all incoming waves $\sum \Delta M_x$.

The rigorous simulations of the MCNN performance with a large number of cells using Eqs.(2-4) is a challenging and time-consuming task. The aim of this work is to show the potential application of the magnetic network with spin wave bus for image processing and to discuss most important advantages and shortcomings. In order to illustrate possible image processing functions, we use a simplified mathematical model for the cell state and an empirical solution of the Landau-Lifshitz equation. The simplified state equation for ME cell at the moment of switching ($V=0$, $M_{th}=0$) can be written in the following form:

$$x_{ij} = f( \sum_{N_r(i,j)} \Delta M_x / M_0 + H_x / H_0 ), \qquad (5)$$

where $x$ is the state variable corresponding to the normalized magnetization projection along the X-axis, $i$ and $j$ subscripts depict cell position in X-Y plane, $f$ is the same piecewise function as in Eq.(1). The argument of the function is the sum of the amplitudes of the incoming spin waves to the cell(i,j) and the x component of the external magnetic field $H_x$. The amplitudes and the external field are expressed in the normalized units, where $M_0$ is the amplitude of the spin wave excited by a single cell, and $H_0$ is the exchange field produced by a single spin wave. From the physical point of view, Eq.(5) the state equation emulates the *H-M* curve for the ME cell in x-projection. The magnetization reversal is by the effective field produced by spin waves and the external field.

In general, the amplitude of the spin wave propagating on the surface of the ferromagnetic film can be described by the wave packet equation for just one magnetization component as follows [21]:

$$\Delta M_x = \frac{C\exp(-t/\tau)}{\delta^4 + \beta^2 t^2} \exp\left[\frac{-\delta^2(r-vt)^2}{4(\delta^4 + \beta^2 t^2)}\right] \times \cos(k_0 r - \omega t + \phi), \qquad (6)$$

where $C$ is a constant proportional to the initial amplitude $M_0$, $\tau$ is the decay time, $r$ is the distance traveled, $k_0$ is the wave vector, $\omega$ is the spin wave frequency, $2/\delta$ is the packet width, $\phi$ is the initial phase, $v = \partial\omega/\partial k \; (k=k_0)$ and $\beta = (1/2)\partial^2\omega/\partial k^2 \; (k=k_0)$ are the coefficients of the first and second order terms, respectively, in the Taylor expansion of the nonlinear dispersion, $\omega(k)$. The initial phase is defined by the cell polarization at the moment of wave excitation: $\phi = 0$ if $M_x > 0$, and $\phi = \pi$ if $M_x < 0$. The use of Eq.(6) significantly simplifies calculations, allowing us to find spin wave amplitudes at the point of interest.

The interaction among the ME cells in the described network is not direct but via the spin waves. The information reflecting ME state is in the *phase* of the spin wave. Assuming all ME cells in the array to be identical in size, the spin waves excited by the cells have the same amplitude and frequency and may interfere in the constructive or destructive manner depending on the relative phase at the point of intersection. In turn, the relative phase is defined by the initial phases, the distance traveled to the point of intersection, and the spin wave group velocity. The accumulated phase shift during the spin wave propagation is given by:

$$\Delta\phi = \int_0^r k(\vec{H}_b) dr , \qquad (7)$$

where particular form of the *k(H)* dependence vary for magnetic materials, film dimensions, and the mutual direction of wave propagation and the external magnetic field [26]. For example, spin waves propagating perpendicular to the external magnetic field (surface wave magnetostatic spin wave – MSSW) and spin waves propagating parallel to the direction of the external field (backward volume magnetostatic spin wave – BVMSW) may obtain significantly different phase shifts for the same field strength. The phase shift $\Delta\phi$ produced by the external magnetic field variation $\delta H$ in the ferromagnetic film can be expressed as follows [27]:

$$\frac{\Delta\phi}{\delta H} = \frac{l}{d}\frac{(\gamma H)^2 + \omega^2}{2\pi\gamma^2 M_s H^2} \text{ (BVMSW)}, \quad \frac{\Delta\phi}{\delta H} = -\frac{l}{d}\frac{\gamma^2(H + 2\pi M_s)}{\omega^2 - \gamma^2 H(H + 4\pi M_s)}\text{(MSSW)}, \qquad (8)$$

where $\Delta\phi$ is the phase shift produced by the change of the external magnetic field $\delta H$, *l* is the propagation length, *d* is the thickness of the ferromagnetic film. Increasing or decreasing the bias field, it is possible to modulate the phases of the propagating spin waves. This fact gives us an intrigue possibility to control the cell-to-cell communication via the external global parameter – magnetic field. In the rest of this Section, we present the examples of different logic functions accomplished on the same template by varying only one global parameters- the bias magnetic field $\vec{H}_b$.

In Fig.7, we show the results of the numerical modeling using Eq.(5-6) to illustrate the image processing by MCNN. The mesh consists of 100×51 magneto-electric cells. The black marks depict ME cells polarized along the X axis. ME cells polarized opposite to X axis are shown in white color. The input image is shown in the top left corner in Fig.7. The first image processing function we apply is filtering, where each cell changes its state to the absolute majority of the neighbor cells. For this particular operation, we chose the following bias field $\vec{H}_b$: $H_x=0$, $H_y=0$, and $\Delta\phi(H_z)=0$. We assume to adjust magnetic filed in Z-direction so that $\Delta\phi=0$. As the bias field has zero projection along the X axis, the cells have no preferable polarization (along or opposite the X axis) and change their state according to the majority of the incoming spin waves. Several images in Fig.7 illustrate the input image transformation after the subsequent steps of filtering.

Next, we show the examples of dilation and erosion functions in Fig.8. To accomplish dilation function, the cell should have the preferable direction for switching along the X-axis. The latter can be easily achieved by introducing a positive x-component of the external field: $H_x/H_0=3$, $H_y=0$, and $\Delta\phi(H_z)=0$. In this case, the ME cell becomes polarized along the X axis if just one of its neighbor cells is polarized along the X axis. Erosion is the function reverse to dilation. The cell should have the preferable direction of switching opposite to the X-axis. For erosion, the components of the external magnetic field are the following: $H_x/H_0=-3$, $H_y=0$, and $\Delta\phi(H_z)=0$. The ME cell becomes polarized opposite to the X axis if just one of its neighbor cells is polarized opposite to the X axis. The pictures in Fig.8 show image evolution after two steps of dilation and two steps of erosion, respectively. The input image is shown in Fig.7(D).

A little bit more sophisticated are the conditions leading to the Horizontal Line Detection (HLD) and the Vertical Line Detection (VLD). In order to accomplish these functions, the cells should have different strength of coupling with their horizontal or vertical neighbors. With spin waves, this task can be accomplished by introducing different group velocities for spin wave propagation along the X and Y axes. The difference in the propagation speed will affect the amplitudes of the transmitted waves as it can be seen from Eq.(6). For example, taking $v_x=r_0/t$ and $v_y=2r_0/t$ (where $r_0$ is the inter-cell distance), the amplitude of the spin wave incoming in Y-direction is $e$ times less than the amplitude of the wave incoming in X-direction at the time of switching. In Fig.9, we present the results of modeling showing HLD and VLD functions. We assumed $H_x/H_0=-1$, $H_z=0$, $H_y$ is adjusted to provide $v_x/v_y=10$ and $v_x/v_y=0.1$ for HLD and VLD, respectively. The pictures in Fig.9 show the results of image processing after three steps of HLD followed and after two steps of VLD, respectively.

Another processing functions, which can be realized in MCNN by manipulating the external magnetic field, are image inversion and edge detection. Inversion is an operation where each pixel changes its state to the opposite. Black pixels become white and vice versa. This operation can be accomplished by providing a π-phase shift for the spin wave during the cell-to-cell propagation: $H_x=0$, $H_y=0$, and $\Delta\phi(H_z)=\pi$. The result of image inversion is shown in Fig.10(A). The edge detection function requires the switching of black pixels, which are surrounded by all black pixels. The preferable switching in this case is achieved by taking $H_x/H_0=-3$, $H_y=0$, and $\Delta\phi(H_z)=\pi$. The picture in Fig.10(B) shows the results of edge detection accomplished for the input image shown in Fig.7(D). Presented results obtained by numerical modeling illustrate the potential of using MCNN for image processing. In principle, any sequence of the processing functions illustrated above can be realized on the same magnetic template *without rewiring or reconfiguration*.

IV Discussion

The described MCNN is based on the combination of two major elements: multiferroic cell and spin wave bus. In our preceding works, we analyzed the efficiency of the spin wave bus [28], and presented the results of numerical modeling illustrating the operation of the ME cell [15, 24]. Here, we would like to use some of the estimates to evaluate the performance c haracteristics of the network comprising a large number of cells connected by the spin wave bus. The scalability of the described MCNN is limited by several factors: the minimum size of the ME cell, the speed of the ME cell switching, and the network tolerance to structure defects. Regular nano-meter size templates can be fabricated by the nanoimprinting technique [29], which makes possible to achieve cell density up to $10^{10}$ cm$^{-2}$. At the same time, the smaller cell size will require more precise control (order of picoseconds) of the switching time sequence during the read-in and read-out procedures as described in the Section II. Another important parameter limiting the scalability of MCNN is the ratio between the operating wavelength and the size of the ME cell. The inevitable variations of the ME cell size will affect the amplitudes of the spin waves excited by the cells. The operation of the network is based on the superposition of the spin waves excited by the nearest neighbor cells (five waves interference). In our numerical simulations, we assumed the amplitudes of the spin waves excited by the ME cells in the array to be the same. The results of the numerical modeling will be the same if we introduce up to 20% variation of the wave amplitudes. The experimentally realized spin-wave based logic devices [19, 27, 30] utilizing two-wave interference show pretty good noise-to-signal ratio (1:10) allowing the recognition of the in-phase and out-of phase outputs. The observed good noise-to-signal ratio is partially due to the long (micrometers) wavelength, which makes the devices immune to the structure imperfections which characteristic size is much less than the wavelength. The shorter is the wavelength the more sensitive is the amplitude of the spin waves to the material imperfections. Taking 100nm as a reasonable benchmark for the operating wavelength, we estimate the practically feasible size of the ME cell as 100nm×100nm and the inter-cell distance 100nm, which give us the cell density of $10^9$ cm$^{-2}$.

According to the model for a single magneto-electric cell presented in [24], we estimate the energy consumed by a single ME cell per switch to be about $10^4 kT$. For this estimate, we used the following data: cell area 100nm×100nm, piezoelectric layer thickness 50nm, and dielectric constant $\varepsilon$=1000, magneto-electric coupling constant 100mV/(cm·Oe), and switching field 200Oe. By multiplying the energy per cell per switch on the cell density, and assuming 1GHz operation frequency, we obtain the consumed power of 40W/cm$^2$. Theoretically, the energy per switch for the magneto-electric cell can be reduced by scaling down the area of the cell, and/or by using more efficient magnetic/piezoelectric pairs. It may be possible to use ferromagnetic material with two easy-axes to avoid the static power consumption for maintaining the cells in the certain polarized state. An example of a voltage control of magnetization easy-axes for spin switching in ultrahigh-density *nonvolatile* magnetic random access memory has been recently experimentally demonstrated [31]. In the best scenario, the power consumed by the ME cells can be decreased by one or two orders of magnitude.

The total power consumed by the MCNN includes also the power to excite spin wave during the read-in process, the power consumed for read-out, and the power to generate the external magnetic field. The problem of energy efficient magnetic field generation is inherent for all proposed spin-based logic schemes [9, 11, 32]. It may be quite possible that the main power consumption will occur in the external contours producing magnetic field rather than inside the magnetic logic devices. At this moment, we would like to exclude this problem from the consideration, as the most important parameter is not the total consumed power but the power dissipated inside the spin wave bus, which may result in the temperature increase and magnetic properties degradation. Another source of power dissipation inside the spin wave bus is the damping of the spin waves excited by the edge ACPS lines during the read-in and read-out procedures. In principle, the mechanism of power dissipation is the same as for the ME cells. The amplitude of the spin wave decreases during the propagation due to the magnon-magnon, magnon-phonon and other scattering events. Eventually, the energy losses due to the scattering will be transformed into the thermal energy of the spin wave bus. The ratio of the power dissipated during read-in and the power consumed during the data processing depends on the number of cells in the array $N$, the efficiency of spin wave excitation, and the number of functions performed during the processing. The energy dissipated by the ME cells scales proportional to $N$. By using the wave-interference approach described in Section II, two spin waves are used to read-in/read-out all cells lying on the same diagonal. The energy required for read-in process scales proportional to $\sqrt{N}$. The same trend takes place for the read-out procedure. Assuming the MCNN to be consisted of the dense arrays of ME cells and performing at least several processing functions on one input data, the energy dissipated by the ME cells will be dominant. Thus, we take 40W/cm$^2$ as a benchmark for power dissipation inside the network, which does not include the power dissipated in the external circuit for magnetic field generation.

The delay time per function in the described network is defined by two factors: spin wave group velocity and the maximum frequency of the ME cell switching. The typical speed of propagation of magnetostatic spin waves in conducting ferromagnetic materials (e.g. NiFe, CoFe) is about $10^6$cm/s. The speed of spin wave propagation is mainly defined by the material properties and cannot be drastically increased. Though the time delay for inter-cell propagation is only 0.1ns-1ps, the time required for spin wave to travel across the network increases proportional to $\sqrt{N}$. In order words, the read-in procedure may take longer time than the computation itself. It may be possible to speed-up the read-in and read-out procedures by introducing additional excitation/detection ports on the edge of the structure. At some point, the time delay on the read-in procedure is the overhead for the structure simplicity and the convenience of having just two input/output ports. The second factor contributing to the delay is the limited frequency of the ME cell switching. The lack of experimental data does not allow us to conclude on the maximum switching frequency for the ME cell. A high frequency (1GHz) PZT nanopowder based oscillators have been demonstrated [33]. At this moment, it is not clear if the higher operation frequency switching is possible and/or energetically efficient for two-phase multiferroic structures. As a possible alternative to the multiferroic structure, we consider spin torque oscillators (STOs) [34, 35]. It was experimentally demonstrated that the magnetization oscillations induced by spin-transfer in two 80-nm-diameter giant-magnetoresistance point contacts in close proximity to each other can phase-lock into a single resonance over a frequency range from approximately 10GHz to 24 GHz for contact spacing of less than about 200 nm [36]. The switching time can be accomplished in the GHz frequency range. On the other hand, the operation of the spin torque oscillator requires DC bias current. The reported current density is about $10^7 A/cm^2$ [35, 37], which makes the spin torque oscillator-based network energetically hungry.

Finally, we want to compare the projected performance characteristics of the spin-wave MCNN with some of the other nano-architectures such as CMOL-CrossNet, QCA, MQCA, and Dot Image Processor. Each of these architectures possesses unique advantages and original technological solutions. The comparison can be done in terms of cell density, operational frequency, power dissipation and the number of functions which can be realized on one template. One of the promising routs to the future nano-architectures is the integration of the existing CMOS-based circuit with nano-devices. An example of this approach is CMOL- a hybrid architecture combining the advantages of CMOS technology with high-density molecular-scale two-terminal nanodevices [38]. One of the possible CMOL architectures is a neuromorphic network called CrossNet [39]. An elementary cell in CorssNet consists of CMOSs connected with molecular latches via a nanowire crossbar [40]. The estimated density of the elementary cell is about $10^7/cm^2$, and the average cell- to-cell communication delay may be as low as 10 ns [38]. The main advantage of the CMOL-CrossNet is high functional throughput. The network can be programmed or trained to perform a variety of tasks including image recognition. In its mathematical representation, CrossNet is close to the classical CNN model described by Leon Chua [1]. The price of this multi-functional performance is in the relatively low cell density and high power consumption estimated to be about $200W/cm^2$ [38].

Quantum Cellular Automata (QCA) is another approach to nano-architectures [32]. QCA is a transistorless computing paradigm that exploits local interaction among the near-neighbor cells. An elementary cell may be a system of four quantum dots with two confined electrons [32], or a single nanomagnet [9]. For the charge-based QCA, the communication among the near-neighbor cells is via the Coulomb interaction. And for the magnetic QCA (MQCA), the communication is via the dipole-dipole interaction. The philosophy of QCA-MQCA approach is two scale-down the size of the elementary cell to a few quantum dots or a nano-magnet, and to utilize local interaction in order to avoid the use of electric currents. Potentially, this approach may lead to an extremely high cell density (up to $10^{13}/cm^2$) and low power consumption (~ $1W/cm^2$) [41]. On the other hand, such a network with transistorless nano-cells would have restricted functionality - one function per fabricated template. Size variability is another critical issue for QCA, as just one "bad" cell can destroy the performance of the whole network.

A quantum dot image processor (QD processor) is a nano-architecture specially designed for image processing [6]. The processor consists of a 2-D periodic array of metallic islands fabricated on the resonant-tunneling substrate. The islands are resistively and capacitively coupled between the nearest neighbors and the conducting substrate. Designed for image processing applications, an elementary cell in the network is not a single dot but a cluster or a "superdot" comprising about 6400 dots, which can act as a single node interacting with the light wave [6]. The advantage of this architecture is the ability to receive all input data in parallel. Thus, the performing time does not depend on the size of the input image. Another important advantage of the QD processor is in the high defect tolerance, which permits significant single island size variation without affecting the network performance. The price for the parallel performance and defect immunity is in the size of the "superdot", which should be comparable with the optical wavelength. The projected "superdot" cell density is $10^6/cm^2$ with the power consumption as low as $0.1mW/cm^2$ [6].

In Table I, we have summarized some main physical parameters and numerical estimates on the projected performance characteristics of CMOL-CrossNet, QCA and MQCA, QD processor, and MCNN. There is an obvious tradeoff between the complexity of the elementary cell and the functional throughput. The more elements are in the cell the more useful operations can be accomplished on one template. The speed of operation and the power consumption are mainly defined by the physical mechanism used for cell-to-cell communication. In the ultimate limit of QCA approach utilizing the nearest-neighbor interactions only, the total power consumption can be reduced to W/cm$^2$, which is two orders of magnitude lower than for today transistor-based processors. Even lower power consumption is estimated for the QD processor 0.1mW/cm$^2$, but for less dense array [6]. The relative advantages of the described MCNN are structure simplicity and functionality. Though the elementary cell consists of a single ME cell (cell density order of 10$^9$/cm$^2$), it is possible to realize several image processing functions (e.g. dilation, erosion, HLD, VLD, inversion, edge detection) by using the external magnetic field as a global control parameter. The power dissipation inside the spin wave bus is conservatively estimated of the order of 40W/cm$^2$ allowing further reduction depending on the progress in the multiferroic structures development.

There is a list of critical comments and open questions on the described MCNN. (i) The combination of the multiferroic cell with spin wave bus has not been experimentally demonstrated. There are different requirements for the magnetic materials to be used in the spin wave bus and in the ME cell. The material of the spin wave bus should be chosen to transmit spin waves with highest possible speed and with minimum losses. The material of the ME cell have to provide the maximum magnetoelectric coupling. At his moment, it is not clear if it is possible to use the same material for spin wave bus and for the ME cell. In the case of using different materials, there will be a problem of matching the spin-wave impedances for two materials. (ii) The maximum practically achievable speed of the ME cell switching is another unknown parameter. Though the frequency of the spin waves can be hundreds of GHz, there is no guarantee that the ME cells will be able to sustain such a high switching speed (high frequency easy-axis rotation). (iii) In Section II, we suggested to detect the reflected spin waves from an individual ME cell for read-out. How big can be the amplitude of the reflected wave to be detected by the inductive-voltage measurements? This is a question, which deserves special and detailed investigation. (iv) Some of the image processing functions (e.g. image inversion, edge detection) requires a π-phase shift for the propagating spin waves. It might need a strong magnetic field to provide such a phase shift on a short (100nm) inter-cell distance. (v) The permissible spin wave amplitude variation is estimated to be 20%. However, the latter does not imply 20% permissible variation of the cell structure. The variation of the ME cell composition may lead to the non-uniform stress and, as a result, to a not controllable easy-axis rotation, which will significantly affect the amplitude of the excited spin wave. How big are the permissible variations of the ME cell size? This is another question to be clarified. Nevertheless these critical comments, the described MCNN possesses certain advantages and offers an original way of using magnetic nanostructures connected via spin waves. The later makes possible to exploit wave phenomena for achieving parallel data processing, and reduce the number of input/output ports. In the present work, we described the basic principle of operation and outlined the potential advantages of magnetic CNN with spin wave bus.

**Conclusions**

We have described a magnetic cellular network consisting of magneto-electric cells united by the common ferromagnetic film - spin wave bus. The utilization of a continuous film as an information bus suggests a solution to the interconnect problem. There is no individual contacts, wires, or a crossbar structure for cell interconnection. We described an original mechanism of using wave interference for cells addressing in nanometer scale arrays. It may be possible to accomplish read-in and read-out procedures for dense array of nano-cells by using just two microstructures (e.g. high-frequency transmission lines). Another distinct property of the described scheme is the ability to govern the network operation by the external parameter – magnetic field. By manipulating the strength and the direction of the external field, it is possible to realize different logic functions. The examples of image processing have been illustrated by numerical modeling. A set of image processing functions such as filtering, dilation, erosion, vertical and horizontal line detection can be realized on one template without rewiring or reconfiguration. According to the estimates, magnetic CNN with spin wave bus may have $10^9$ cells per $cm^2$ cell density and accomplish data processing in a parallel manner with GHz frequency. The main disadvantage of the proposed scheme is the relatively long time delay required for read-in and read-out procedures. The time delay is due to the limited spin wave velocity and due to the sequential mechanism for data loading and retrieval. The lack of experimental data does not allow us to conclude on the practically achievable minimum switching energy per ME cell and the maximum switching frequency. The realization of the magnetic CNN with spin wave bus would require an extensive theoretical and experimental study and may provide a route to logic circuits for special task data processing.

Acknowledgments

The work was supported in part by the Focus Center Research Program (FCRP) Center of Functional Engineered Nano Architectonics (FENA) and by the Nanoelectronics Research Initiative (NRI) - Western Institute of Nanoelectronics (WIN).

Figure captions

Fig.1 Schematics of MCNN in conjunction with electronic circuits. One electronic circuit is to provide the input data, and the second circuit is to define the data processing function. The input data are received in the form of voltage pulses and converted into spin wave signals by the ACPS lines. The information exchange inside MCNN is via spin waves only. The core of the MCNN is a layered structure comprising from the bottom to the top a semiconductor substrate (e.g. silicon), a conducting ferromagnetic layer (e.g. NiFe), an insulating layer (e.g. silicon dioxide) with a regular two dimensional array of nano-patterns filled with a non-conducting piezoelectric material (e.g. PZT), and a metallic contact (e.g. Al) on the top. The elementary cell structure is shown on the right. It comprises the ferromagnetic under the nano-pattern, the piezoelectric inside the nano-pattern, and the metal contact on top. An electric field applied across the piezoelectric material produces stress. In turn, the produced stress affects the anisotropy field in the ferromagnetic material and results in the local easy-axis rotation as shown in the right down corner. The ferromagnetic layer is polarized along the Z-axis. The applying of the electric field rotates the easy axis under the piezoelectric nano-patterns from Z-axis toward the X-axis.

Fig.2 Three basic states of the magneto-electric cell. (A) There is no bias voltage applied. And the ferromagnetic material is polarized along the Z axis. (B) Under a non-zero voltage applied, the easy axis of the ferromagnetic layer under the piezoelectric

rotates towards the X axis. The rotation makes two directions of magnetic polarization (along or opposite to the X axis) energetically preferable. The energy barrier ΔE between the two possible polarizations is defined by the angle of the easy-axis rotation. (C) The ultimate limit of 90 degree easy axis rotation. The cell has the biggest $M_x$ component and the highest energetic barrier ΔE.

Fig.3 Illustration of the read-in procedure. Two coherent spin waves are excited by the ACPS lines. The waves propagate along the X and Y axes. (a) Three-dimensional plot showing the magnetization change as a result of the waves interference. (b) The contour plot showing the position of the two waves intersection after the propagation in X-Y plane, where $t_0$ is the time spin wave travels across a single ME cell. (c) Three snapshot illustrating the movement of the point of interference in time. The switching of the cell occurs by decreasing the bias voltage at the moment when the point of interference crosses the cell.

Fig.4 Spin wave exchange among the ME cell. All cells simultaneously emit spin waves as a result of the bias voltage change. The amplitude of the emitted spin waves is the same for all cells, but the phases of the waves depend on the initial cell polarization. The wave propagate in the ferromagnetic film – spin wave bus in X-Y plane. The speed of propagation is controlled by the strength and direction of the external magnetic field.

Fig.5 Illustration of the read-out procedure. The cell in the array is excited by the superposition of two plane waves generated by ACPS lines similar to the read-in procedure. The cell produces scattered spin waves propagating through the spin wave bus. The state of the cell is reflected in the phase of the scattered spin wave, which can be detected by the edge ACPS lines.

Fig.6 The results of numerical modeling showing the magnetization dynamics of a single ME cell. At the initial moment of time, the cell is polarized along the Z-axis ($M_x = M_y = 0$, $M_z = 1$). There is no bias voltage applied and the easy axis is parallel to the Z direction. Next, an incoming spin wave arrives to the ME cell. At the moment of the spin wave arrival, the bias voltage is applied initiating the easy-axis rotation towards the X-axis. There are two evolution curves shown in the figure: the solid curve corresponds to the case when the incoming spin wave has a positive x-component ($\Delta M_x > 0$), and the dashed curve corresponds to the case when the incoming spin wave has a negative x-component ($\Delta M_x < 0$). At the end of the switching, the cell becomes polarized along the X axis if $\Delta M_x > 0$, and opposite to the X axis if $\Delta M_x < 0$.

Fig.7 The results of the numerical modeling illustrating the image processing by MCNN on the mesh consisting of 100×51 magneto-electric cells. The black marks depict ME cells polarized along the X axis. ME cells polarized opposite to X axis are shown in white

color. (A) input image; (B) input image after one step of filtering; (C) after two steps of filtering; (D) after three steps of filtering.

Fig.8 The results of the numerical modeling illustrating image processing with dilation and erosion functions. The input image is shown in Fig.7(D). (A) Image after one step of dilation; (B) after two steps of dilation; (C) after one step of erosion; (D) after two steps of erosion.

Fig.9 Results of the numerical modeling illustrating the Horizontal Line Detection (HLD) and the Vertical Line Detection (VLD). (A) input image; (B) after one step of HLD; (C) after two steps of HLD; (D) after three steps of HLD; (E) The input image after one step of VLD; (F) after two steps of VLD.

Fig.10 Results of numerical modeling illustrating (A)image inversion and (B) edge detection. The input image is show in Fig.6(D).

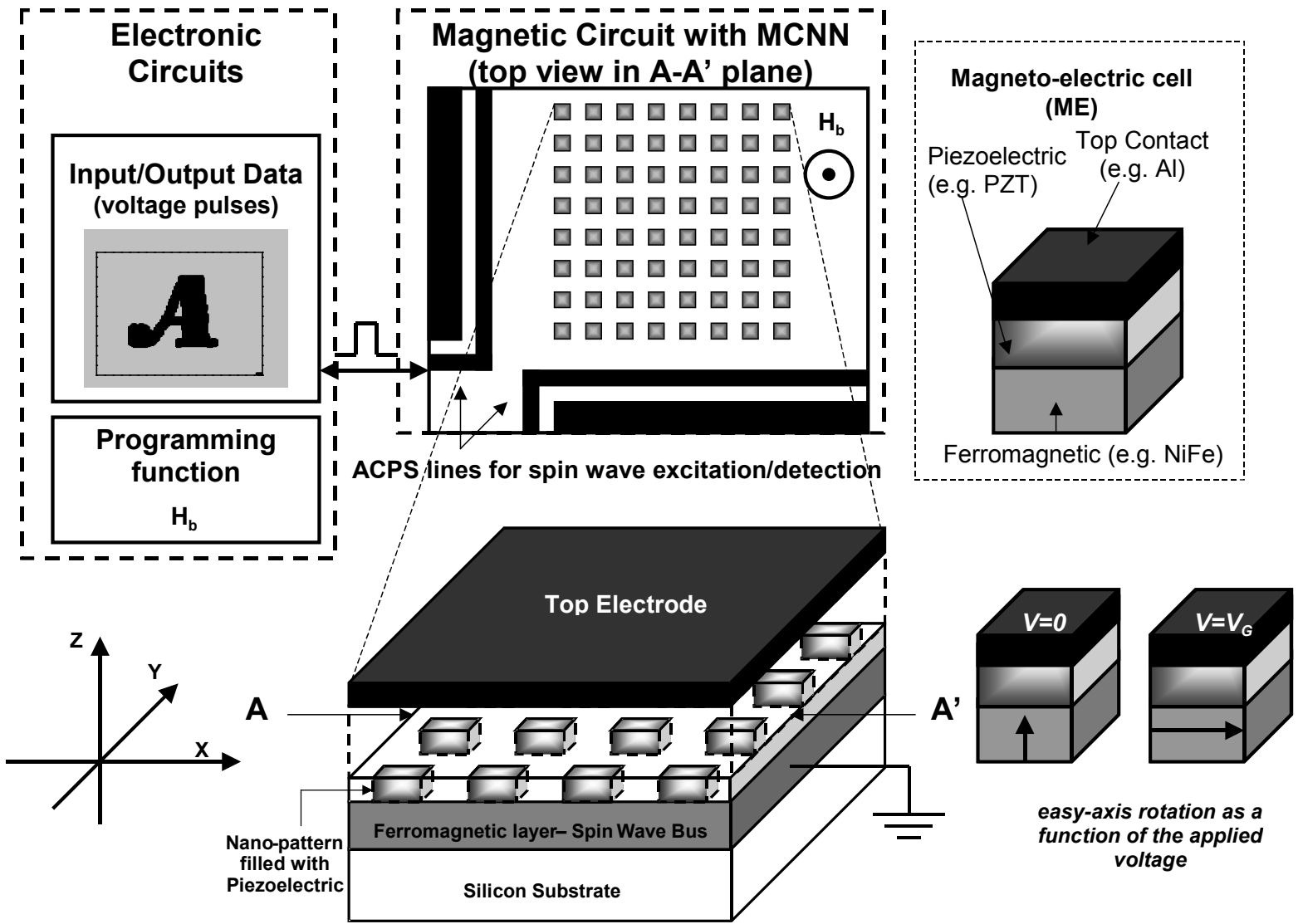

**Fig.1**

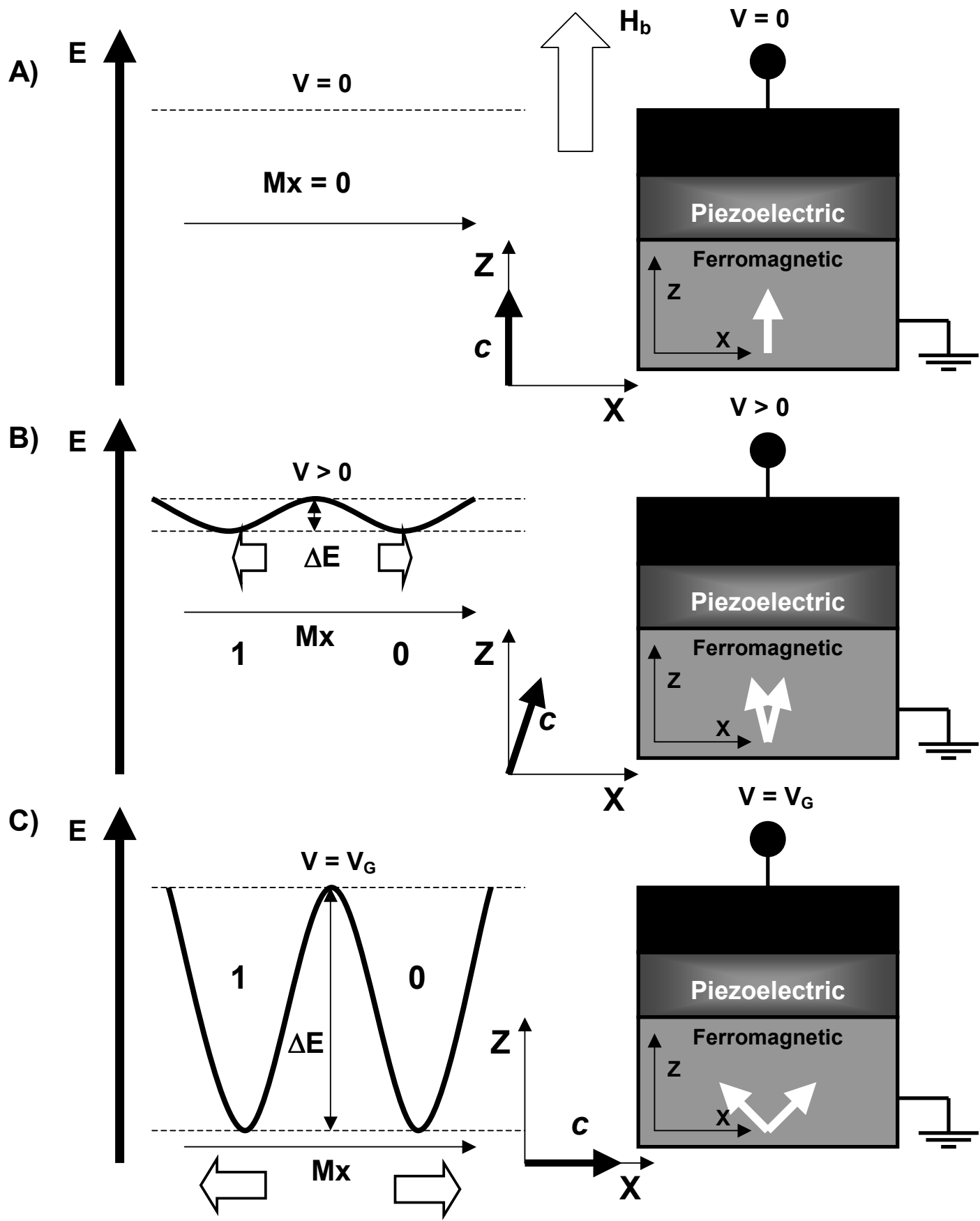

**Fig.2**

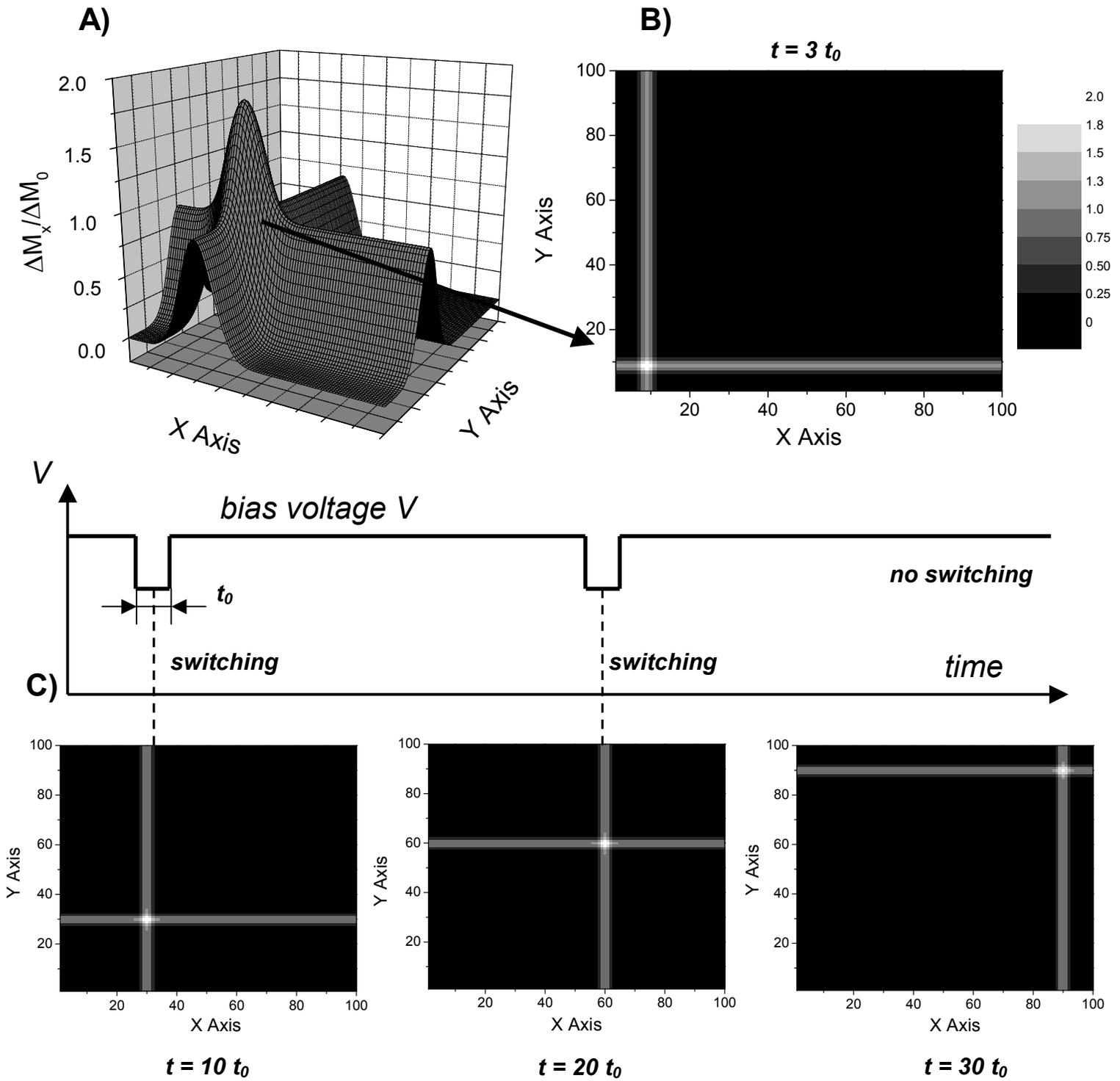

**Fig.3**

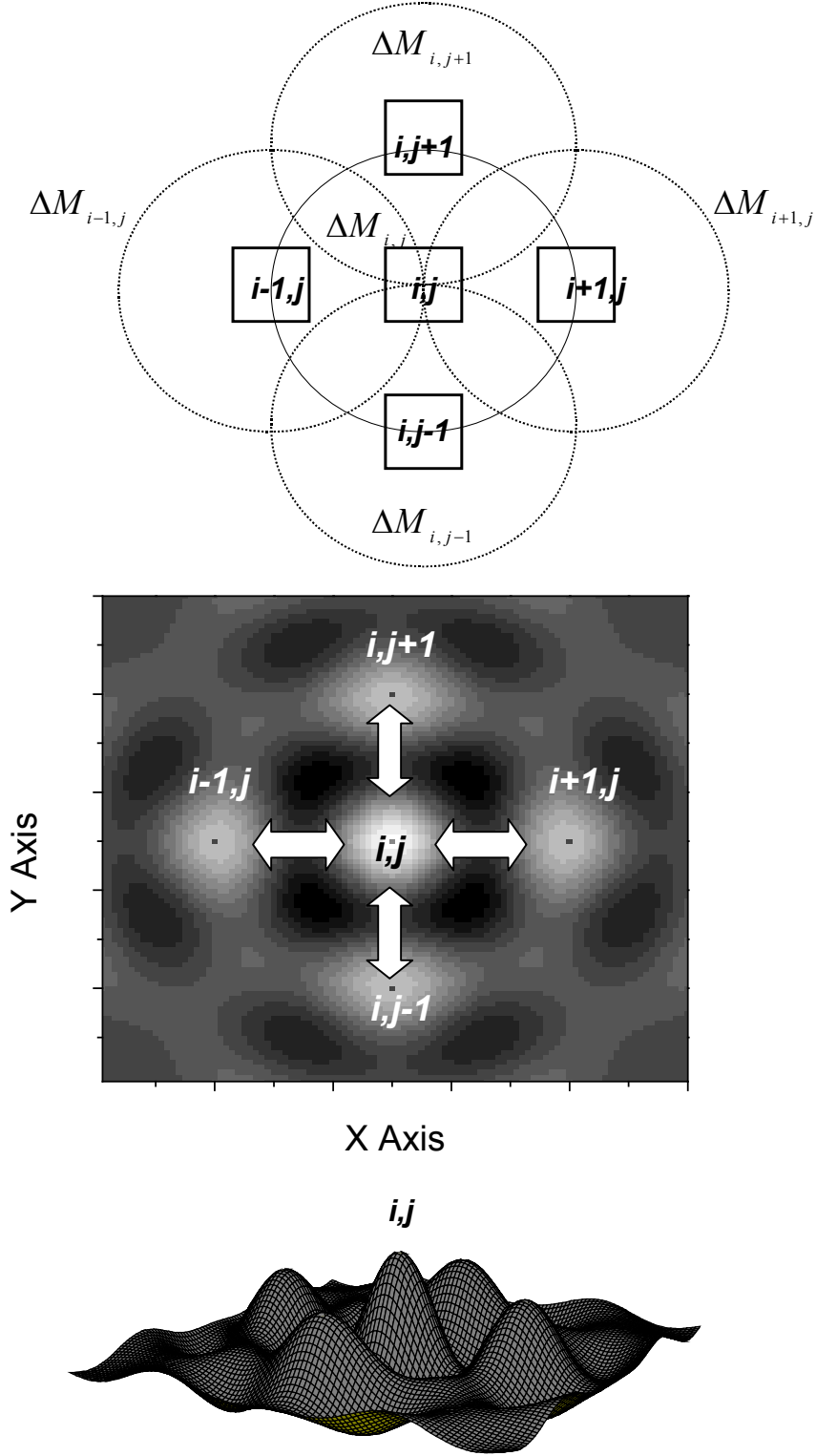

**Fig.4**

*cell excitation by two plane waves*

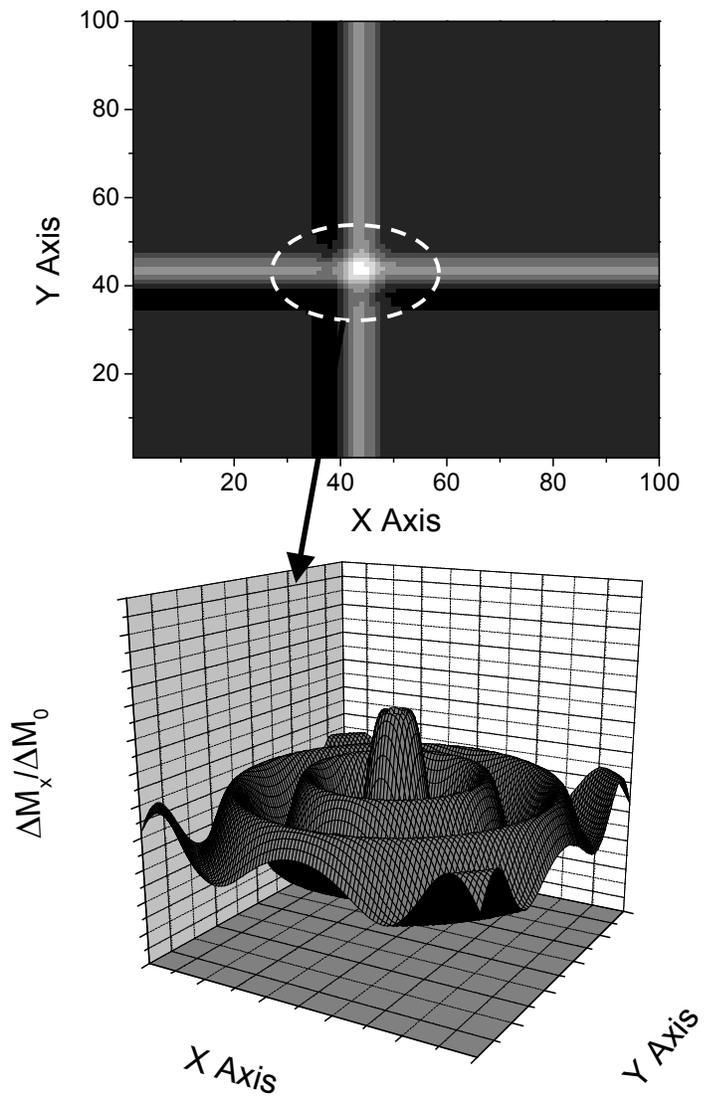

*spin waves emitted by the cell*

**Fig.5**

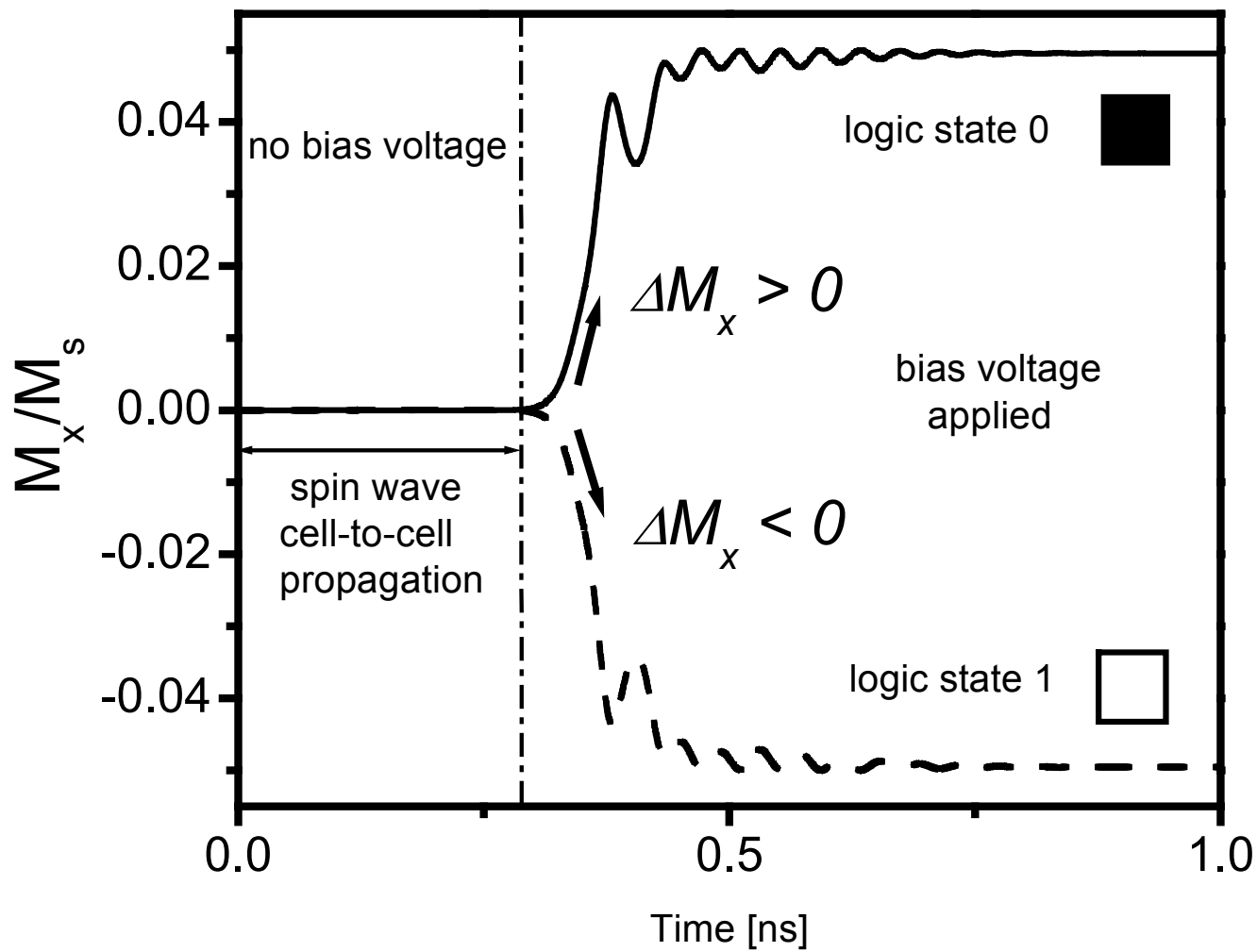

**Fig.6**

*input data*        *after one step of filtering*

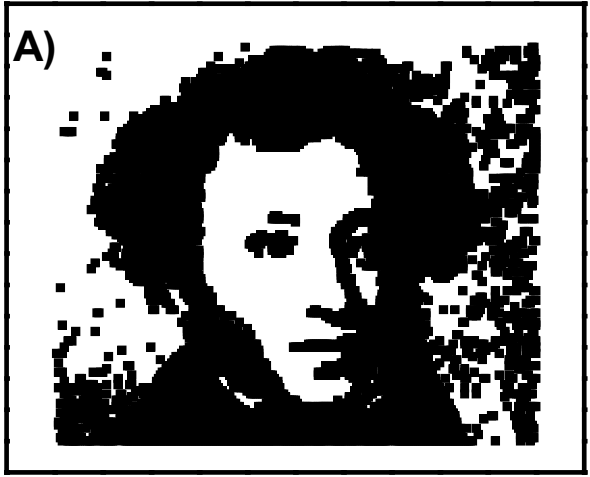 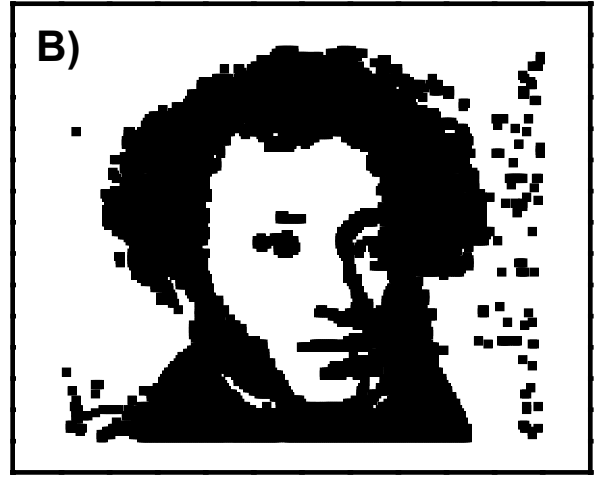

A)        B)

*after two steps of filtering*        *after three steps of filtering*

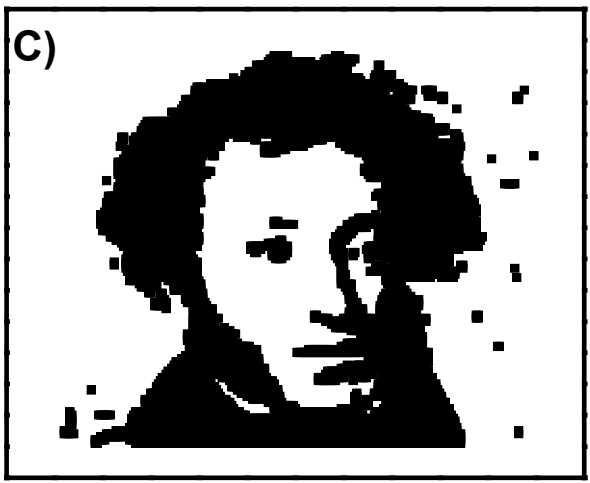 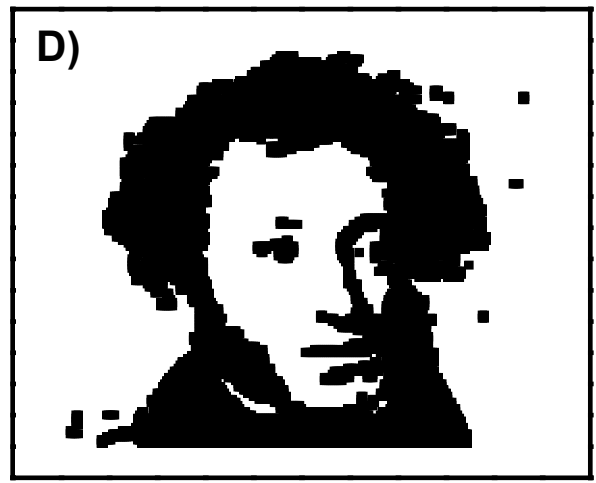

C)        D)

**Fig.7**

*after one step of dilation*   *after two steps of dilation*

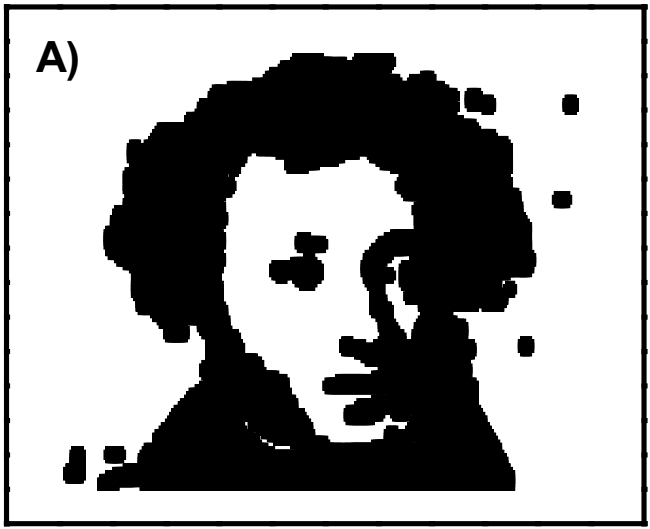 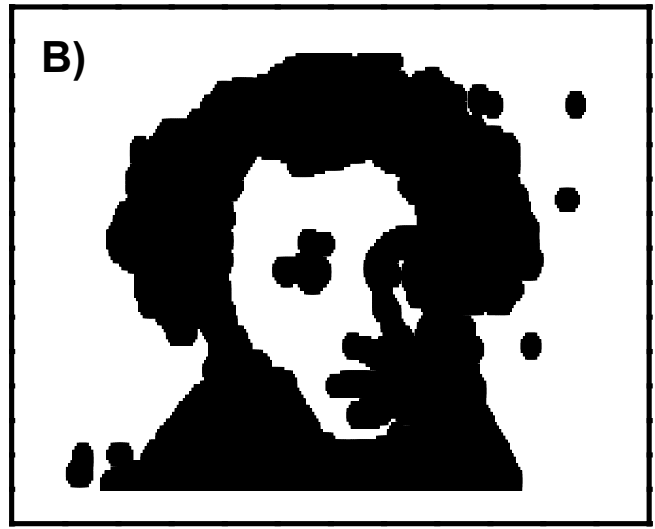

A)   B)

*after one step of erosion*   *after two steps of erosion*

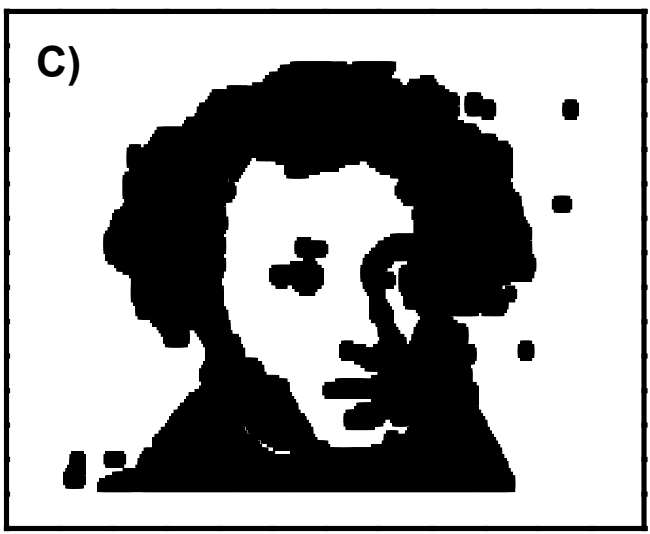 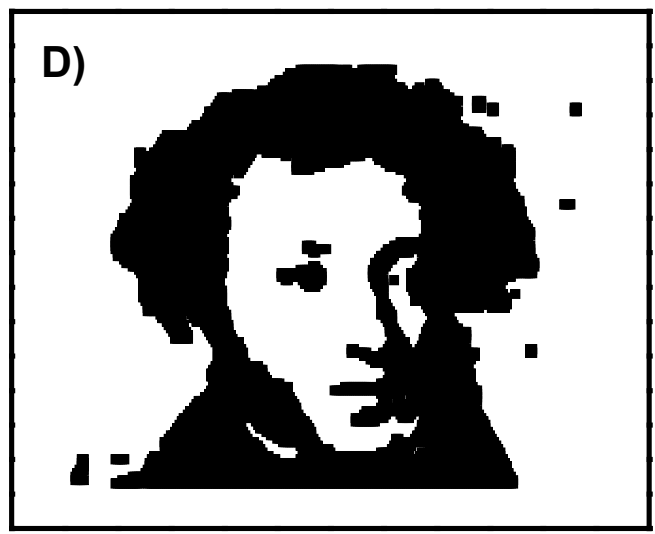

C)   D)

**Fig.8**

*input data*　　　　　　　　*after one step of HLD*

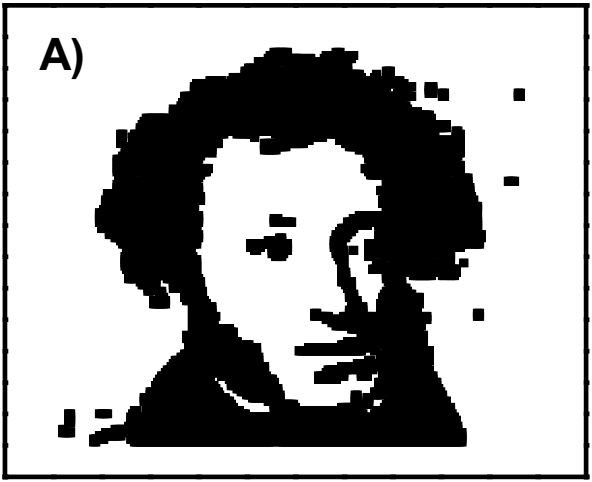 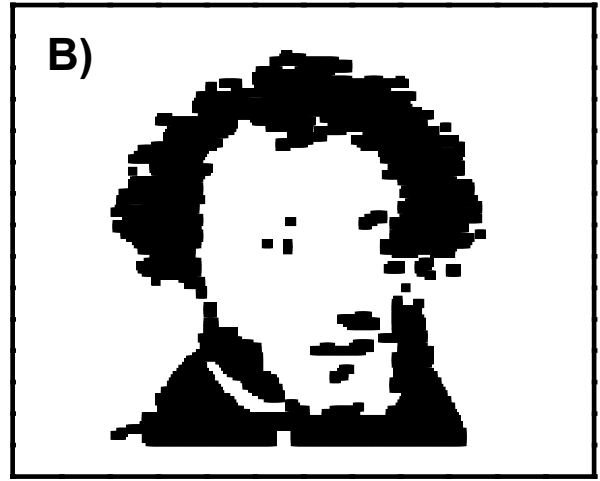

*after two steps of HLD*　　　*after three steps of HLD*

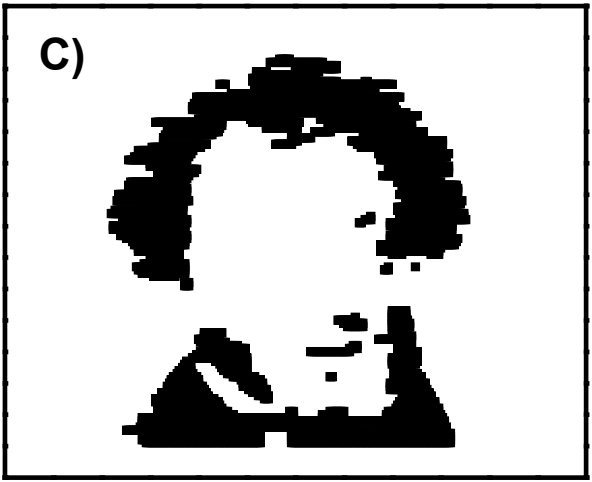 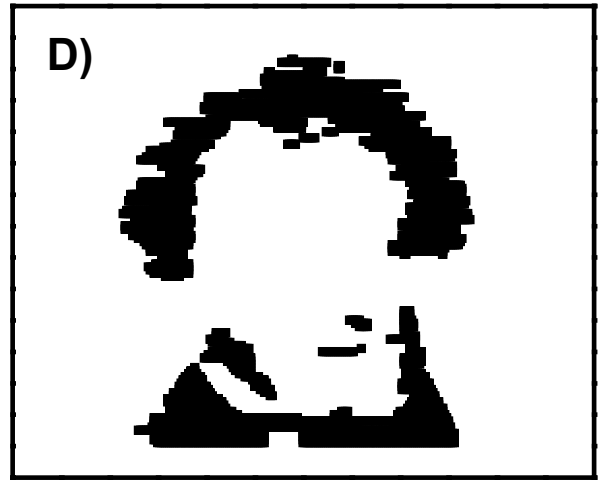

*after one step of VLD*　　　*after two steps of VLD*

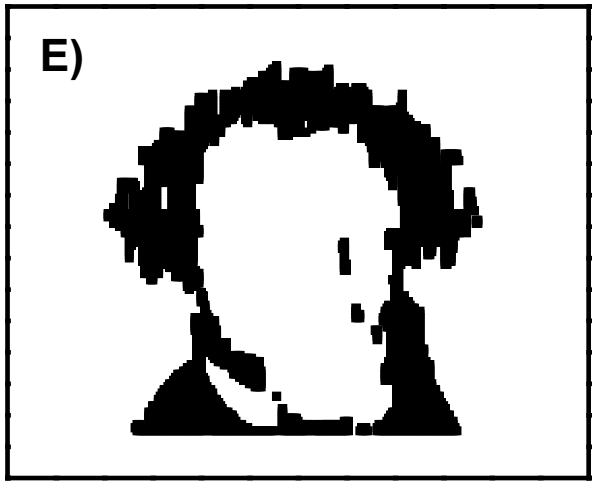 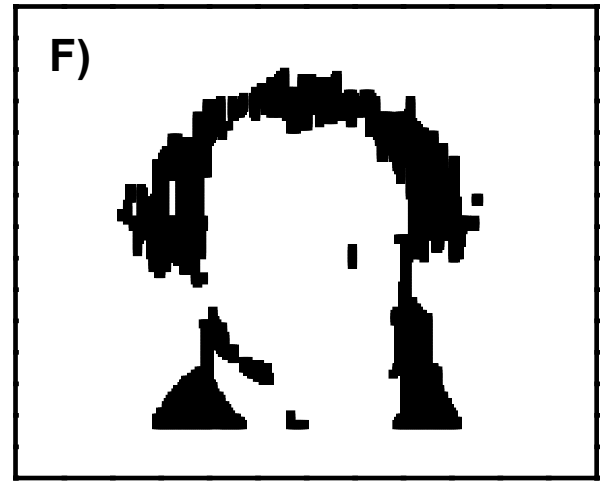

**Fig.9**

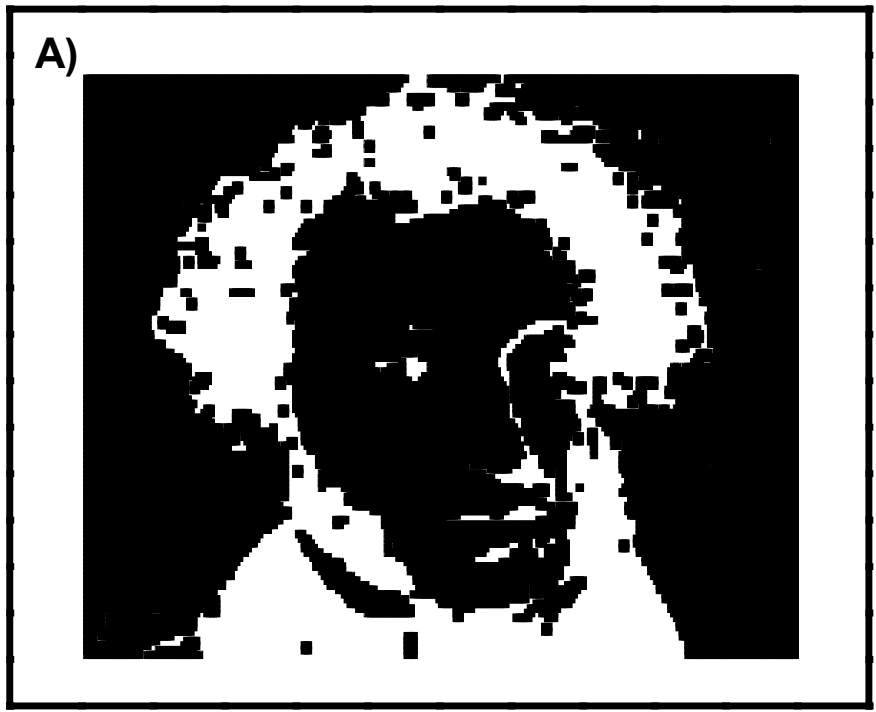
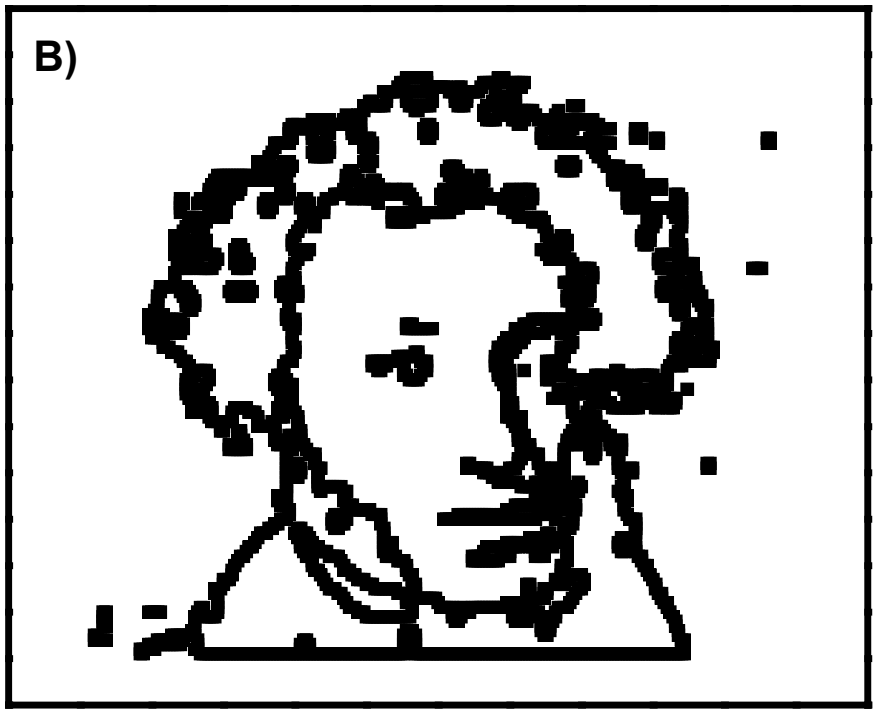

**Fig.10**

|  | **CMOL CrossNet** | **QCA/MQCA** | **QD Image Processor** | **MCNN with Spin Wave Bus** |
|---|---|---|---|---|
| Elementary cell | Molecule latch + CMOS | confined electrons/ nanomagnet | "super" dot | ME cell |
| Cell-to-Cell communication via | electric current | Coulomb interaction/dipole-dipole interaction | capacitive coupling | spin waves |
| Power dissipation mechanism | Joule heat | variability caused losses | Joule heat | spin wave damping |
| Information read-in/read-out | sequential | sequential | parallel | sequential |
| Delay time limiting factor | RC delay | tunneling time/dipole-dipole group velocity | RC delay | spin wave group velocity |
| Functional throughput | reconfigurable-multifunctional | one function per template | one function per template | several functions per template |
| Cell density | $10^7/cm^2$ | $10^{11}/cm^2$-$10^{13}/cm^2$ | $10^6/cm^2$ | $10^9/cm^2$ |
| Operation frequency | 0.1GHz | GHz | GHz | GHz |
| Power dissipation | 200W/$cm^2$ | W/$cm^2$ | 0.1mW/$cm^2$ | 40W/$cm^2$ |